# Thermodynamics-Inspired High-Entropy Oxide Synthesis

Saeed S. I. Almishal[1], Matthew Furst[1], Yueze Tan[1], Jacob T. Sivak[2], Gerald Bejger[3], Dhiya Srikanth[1], Joseph Petruska[1], Christina M. Rost[3], Susan B. Sinnott[1,2,4], Long-Qing Chen[1], and Jon-Paul Maria[1]

[1]*Department of Materials Science and Engineering, The Pennsylvania State University, University Park, PA 16802, USA*

[2]*Department of Chemistry, The Pennsylvania State University, University Park, PA 16802, USA*

[3]*Department of Materials Science and Engineering, Virginia Polytechnic Institute and State University, Blacksburg, VA 24061, USA*

[4]*Institute for Computational and Data Sciences, The Pennsylvania State University, University Park, PA 16802, USA*

**Corresponding Authors:** Saeed S. I. Almishal saeedsialmishal@gmail.com

## Abstract

High-entropy oxide (HEO) thermodynamics transcend temperature-centric approaches, spanning a multidimensional landscape where oxygen chemical potential plays a decisive role. Here, we experimentally demonstrate how controlling the oxygen chemical potential coerces multivalent cations into divalent states in rock salt HEOs. We construct a preferred valence phase diagram based on thermodynamic stability and equilibrium analysis, alongside a high throughput enthalpic stability map derived from atomistic calculations leveraging machine learning interatomic potentials. We identify and synthesize seven equimolar single-phase rock salt compositions that accommodate multivalent Mn, Fe, or both, as confirmed by X-ray diffraction and fluorescence. X-ray absorption fine structure spectra reveal predominantly divalent cations. Ultimately, we introduce oxygen chemical potential overlap as a key complementary descriptor predicting HEO stability and synthesizability. Although we focus on rock salt HEOs, our methods are chemically and structurally agnostic, providing a broadly adaptable framework for navigating HEOs thermodynamics and enabling a broader compositional range with contemporary property interest.

## Introduction

High-entropy oxides (HEOs) redefine ceramics discovery by harnessing chemical disorder to unlock otherwise inaccessible chemistries through enthalpy-minimization approaches. Despite HEOs' growing research body, much remains to be uncovered about the principles governing their stability and formation into single-phase multicomponent materials[1–3]. Addressing these knowledge gaps is essential to advancing HEO discovery and development. Undoubtedly, configurational entropy plays a critical role in stabilizing multi-component solid solutions, especially at elevated temperatures where the thermal energy of mixing ($-T\Delta s_{mix}$; where T is temperature and $\Delta s_{mix}$ is entropy of mixing per mol dominated by configurational entropy) rivals or exceeds the enthalpy of mixing per mol ($\Delta h_{mix}$) in minimizing the solid solution chemical potential ($\Delta\mu = \Delta h_{mix} - T\Delta s_{mix}$)[1,3–5]. However, single-phase stability and synthesizability is not guaranteed by simply increasing configurational entropy; enthalpic contributions and thermodynamic processing conditions must also be carefully considered.

The Hume-Rothery rules offer straightforward yet effective guidelines for predicting solid solution formation based on enthalpic considerations[6]. The prototypical HEO, $Mg_{1/5}Co_{1/5}Ni_{1/5}Cu_{1/5}Zn_{1/5}O$ (MgCoNiCuZnO for brevity) does not comply with the Hume-Rothery crystal structure compatibility criterion, as 2/5 cations prefer alternative crystal structures — ZnO favors the wurtzite structure, while CuO prefers the tenorite structure[1–3]. Therefore, both the high configurational entropy and the rock salt structure inherent stability, which possesses the broadest basin of attraction in binary oxides[7], drive the single-phase rock salt solid solution formation. MgCoNiCuZnO adheres closely, however, to every other Hume-Rothery criteria, including cation radii, electronegativity, and valence compatibility. The largest size disparity occurs between $Ni^{2+}$ and $Co^{2+}$, with $Co^{2+}$ being 8% larger than $Ni^{2+}$ — still within the 15% Hume-Rothery limit. This criterion supports why Ca, Sr, or Ba cannot be incorporated in equimolar amounts into MgCoNiCuZnO under equilibrium synthesis conditions[8]. Finally, MgCoNiCuZnO stability is also intrinsically tied to its cations favoring and maintaining a 2+ oxidation state in their binary oxides over the critical 875-950°C temperature range for phase stabilization under ambient oxygen partial pressure ($pO_2$), along with minimal electronegativity variation among the 2+ cations[9]. This valence compatibility criterion explains why incorporating persistent $Sc^{3+}$ into equimolar rock salt HEOs is challenging, if not impossible, under equilibrium synthesis conditions, despite its ionic radius being comparable to other cations in MgCoNiCuZnO[10].

To expand the rock salt high entropy oxide library, we need to identify cations whose ionic radii closely match MgCoNiCuZnO average radius and that can be coerced to take a 2+ oxidation state. Unlike Sc, all other 3d transition metals, including and beyond those in MgCoNiCuZnO, can adopt a 2+ oxidation state. However, their stability in the 2+ state depends on their electronic configuration as well as the thermodynamic and kinetic processing factors. Among those, Mn is positioned at the 3d-period center with five unpaired electrons, leading to the largest and highest possible oxidation states in the entire period. Therefore, Mn exhibits remarkable versatility, forming diverse oxide structures and phases. Under ambient atmospheric pressure, Mn commonly exists as tetragonal pyrolusite ($MnO_2$)[11,12]. However, at elevated temperatures, such as those used in our HEO synthesis (~460°C and above), Mn transitions to $Mn_2O_3$ typically adopting the bixbyite or corundum structures[11,12] (Supporting Information Note 1). Following Mn in the periodic table is Fe, which is stable as $Fe_2O_3$ in the hematite phase (corundum structure) under ambient conditions – unlike Mn, Fe is largely restricted to 2+ and 3+ oxidation states. Importantly, both Mn and Fe can be readily coerced into the 2+ oxidation state under lab-accessible reducing conditions. In contrast, while Ti, V, and Cr can also adopt 2+ oxidation states, they require extreme reducing conditions compared to all other cations in the 3d period (Supporting Information Note 5).

Therefore, Mn and Fe are compelling candidates for incorporation into rock salt HEOs, as they can adopt a 2+ oxidation state and maintain size compatibility in both their 2+ and 3+ states, with ionic radii deviations within 15% of the rock salt cations in MgCoNiCuZnO. While entropy stabilization is typically explored through cation selection, widely accepted room pressure, and high temperature, we establish oxygen chemical potential as a powerful yet underutilized thermodynamic axis for controlling phase stability. By precisely tuning $pO_2$ during synthesis, we suppress higher oxidation states and promote $Mn^{2+}$ and $Fe^{2+}$ incorporation. To harness this control in the simplest manner, we propose starting with AO oxide mixtures and employing high-temperature synthesis under a controlled, continuous Argon (Ar) flow to maintain low $pO_2$, effectively steering different compositions toward a stable, single-phase rock salt structure.

## Main

### *Rock salt high entropy oxide thermodynamics*

We begin by examining fundamental thermodynamic variables to assess rock salt HEOs' stability and develop intuition into their synthesizability. In Figure 1(a), we present an enthalpic stability map (adapted from our study in Ref.[8]); with mixing enthalpy ($\Delta H_{mix}$) in meV/atom and bond length distribution ($\sigma_{bonds}$) in Å as its axes[8]. $\Delta H_{mix}$ represents the enthalpic barrier to single-phase formation, while $\sigma_{bonds}$ quantifies lattice distortion through the relaxed first-neighbor cation-anion bond lengths standard deviation. A lower $\sigma_{bonds}$ suggests minimal lattice distortion, promoting single-phase stability in accordance with the Hume-Rothery ionic size rule. To construct and populate this stability map, we leverage the Crystal Hamiltonian Graph Neural Network (CHGNet) machine learning interatomic potential, which achieves near-density functional theory accuracy with considerably reduced computational cost[13]. The map includes all equimolar four-, five- and six-component compositions drawn from the cation cohort: Mg, Ca, Co, Mn, Fe, Co, Ni, Cu and Zn (all numerical values can be found in Supporting Information Note 6 Table S2). It stands out that all five-component compositions containing Mn and Fe, but lacking Ca and Cu (highlighted in blue in the map) exhibit the lowest $\Delta H_{mix}$ and $\sigma_{bonds}$ values among all other five-component compositions, with values even lower than the prototypical MgCoNiCuZnO[8,14,15]. Despite these favorable characteristics, Mn and Fe-based compositions have eluded conventional synthesis routes for the past decade[1,15,16]. MgCoNiMnFeO is the only composition among those highlighted in the map that has been experimentally reported[15]. Pu et al. synthesized it in an elaborate method starting from oxalate precursors followed by annealing under a controlled atmosphere that involved both reducing and oxidizing agents[15]. Notably, MgCoNiMnFeO exhibits the lowest $\Delta H_{mix}$ and $\sigma_{bonds}$ among the six five-component compositions that exclude Ca and Cu. This arises because it lacks Zn, which preferentially stabilizes in a wurtzite — rather than a rock salt — structure in its oxide form. Consequently, to our best knowledge, no equilibrium-synthesized rock salt HEOs containing Mn, Fe, and Zn have been reported.

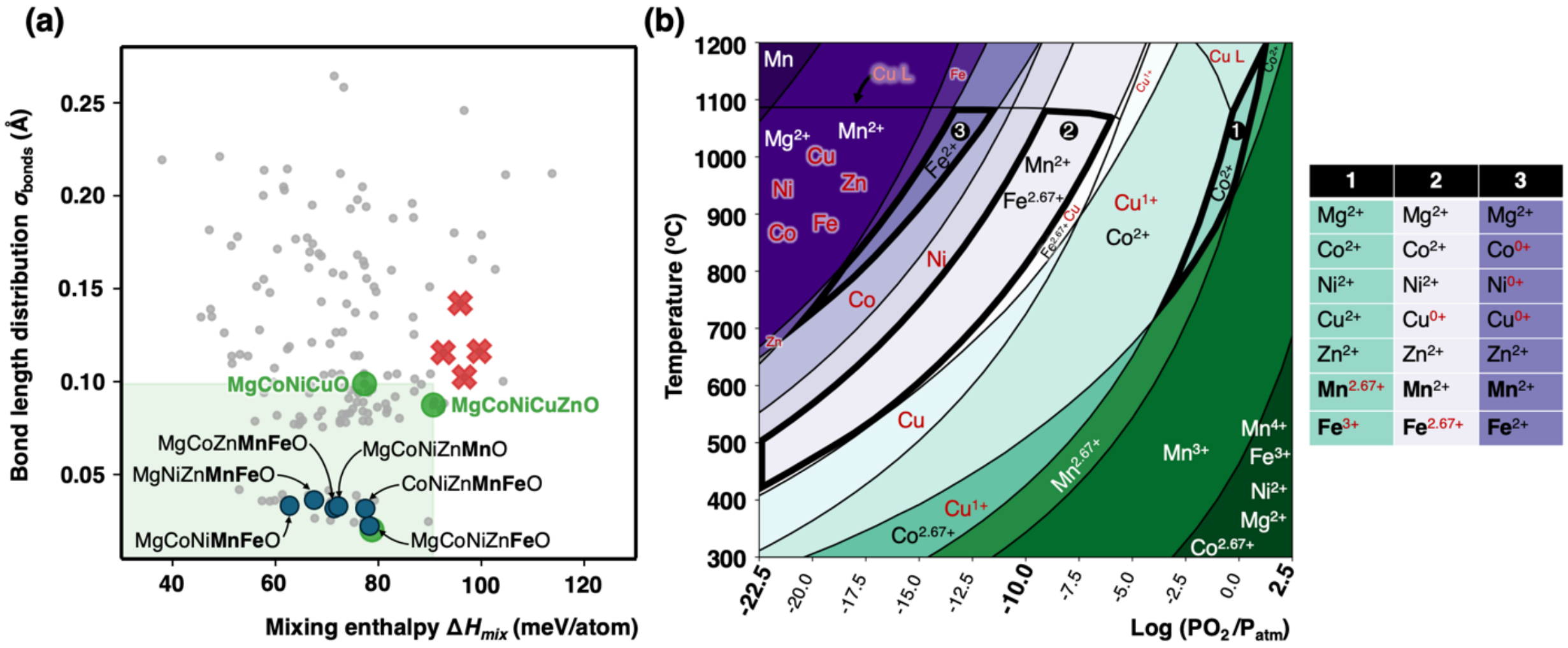

Figure 1: (a) Rock salt HEO composition map with $\Delta H_{mix}$ and $\sigma_{bonds}$ (adapted from our study in Ref.[8]). Experimental synthesis results for single and multi-phase are indicated with green circles and red crosses, respectively. Region predicted as single-phase are shown in a light green shade. All 5-component compositions in this study are labeled and indicated with blue circles. (b) Temperature and oxygen partial pressure phase diagram illustrating each cation preferred oxidation state in its stable binary oxide phase. In region (1) all cations in $Mg_{1/5}Co_{1/5}Ni_{1/5}Cu_{1/5}Zn_{1/5}O$ are stable in 2+ oxidation state in their $A^{2+}O^{2-}$ binary oxide form; in region (2) all cations in $Mg_{1/5}Mn_{1/5}Co_{1/5}Ni_{1/5}Zn_{1/5}O$ are stable in 2+ oxidation state in their $A^{2+}O^{2-}$ binary oxide form; and in region (3) Fe is stable in the 2+ oxidation state in its $A^{2+}O^{2-}$ binary oxide form, while Ni and Zn reduce to their metallic states. Because Ti, V, and Cr require substantially lower oxygen partial pressures to maintain their 2+ oxidation states, we excluded them from this analysis (see Supporting Information Note 5).

To reconcile the low $\Delta H_{mix}$ for Mn and Fe-containing compositions with their synthesis challenges, we use CALPHAD to construct a temperature–oxygen partial pressure phase diagram (Figure 1(b)). This diagram maps the stable Mg and 3d transition metal (Mn through Zn) oxidation states in their binary oxide phases and delineates temperature-pressure zones offering partial or complete overlapping stability. We focus on three distinct regions, each designated by a number and outlined in bold, with the stable valence of each cation presented in the accompanying side table. The phase diagram reveals that under ambient conditions (the far bottom-right region), Mn predominantly adopts a 4+ oxidation state, Fe a 3+ state, Co averages 2.67+ (due to mixed valencies), and the remaining cations persist in the 2+ state. In contrast, at extremely low oxygen partial pressures ($<10^{-15}$ bar) and temperatures above ~800°C, all cations except Mg eventually reduce to their metallic forms. In Region 1 (ambient pressure, T > ~875°C), only the cations in prototypical MgCoNiCuZnO are stable in their $A^{2+}O^{2-}$ binary oxide phases (see side table), explaining this composition's unique stability under ambient conditions. Deviations from Region 1, either toward higher temperatures or lower $pO_2$, lead to CuO reduction and Cu melting. Absent Copper, as $pO_2$ decreases from Region 1, Mn reduces to 2+, marking the transition into Region 2, while further reductions stabilize $Fe^{2+}$, defining Region 3 where, importantly, Mn remains 2+ stable. Therefore, Regions 2 to 3 outline the synthesis conditions under which rock salt high-entropy oxides containing Mn and Fe, but no Cu, can be stabilized based solely on oxidation-state compatibility criteria. We therefore experimentally explore incorporating Mn and Fe within the rock salt high-entropy oxide family by synthesizing the six Mn- and Fe-containing compositions identified in the stability map in Figure 1(a) and maintaining low $pO_2$ to access regions 2 to 3 in Figure 1(b). The six compositions are as follows; first, Mn is introduced into MgCoNiZnO – a four-component derivative of the prototypical MgCoNiCuZnO obtained by removing CuO – forming MgCoNiZnMnO. Next, Mn is replaced with Fe, yielding MgCoNiZnFeO. Finally, both Mn and Fe are incorporated simultaneously, with one of the four cations in MgCoNiZnO systematically removed in each case, resulting in four additional compositions: MgCoNiMnFeO, MgNiZnMnFeO, CoNiZnMnFeO, and MgCoZnMnFeO.

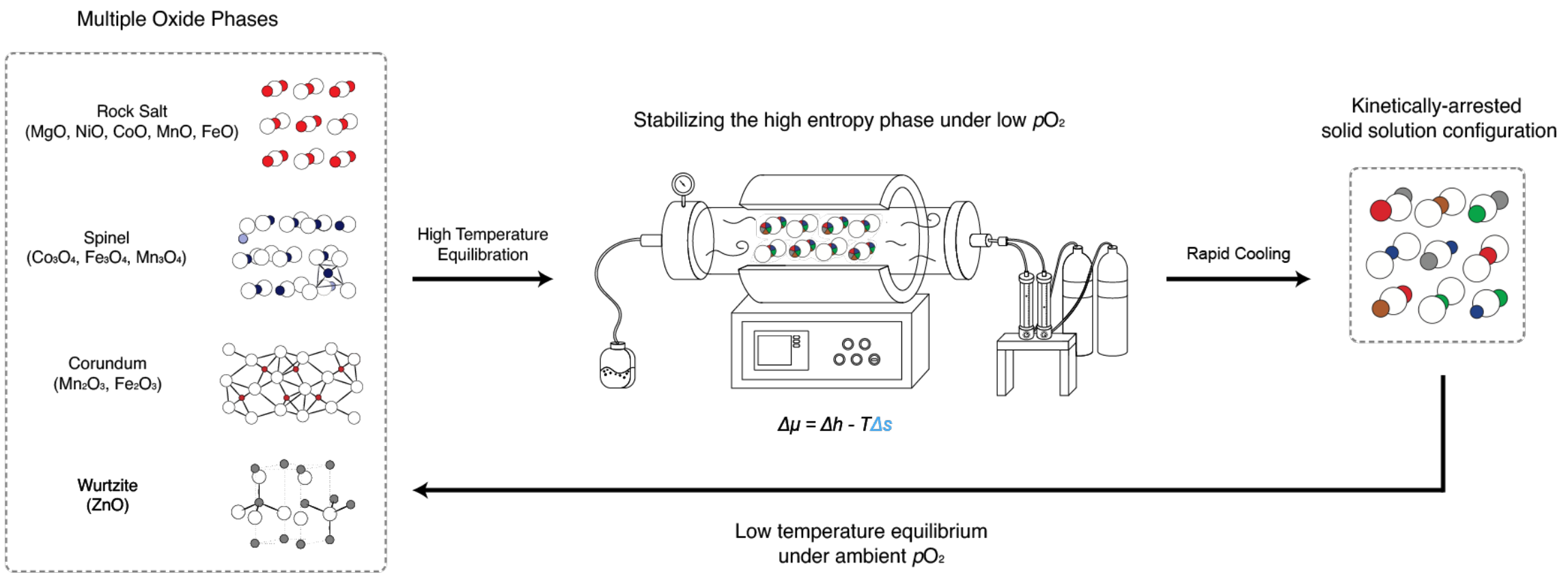

Figure 2: Flowchart illustrating phase progression in the Mn and Fe containing compositions with temperature. We highlight that in addition to wurtzite ZnO, spinel phases ($Co_3O_4$, $Mn_3O_4$ and $Fe_3O_4$) and corundum phases ($Mn_2O_3$ and $Fe_2O_3$) should be explicitly considered in low temperature processes and as competing phases. Note $Mn_2O_3$ can also form bixbyite $Ia\overline{3}$ phase and $Mn_xFe_yO_\delta$ sintered in air forms a spinel and bixbyite mixture, as detailed in Supporting Information Note 1. Δμ is the change in chemical potential, Δh is the change in molar enthalpy, T is temperature, and Δs is the change in molar entropy.

## *Synthesizing Mn- and Fe-containing rock salt high entropy oxides*

We hypothesize all six Mn- and Fe-containing five-component rock salt phases shown in Figure 1(a) can be stabilized by combining $A^{2+}O^{2-}$ binary oxides and processing them at high temperatures under a low $pO_2$ environment. This approach suppresses higher oxidation states, ensuring cation valence states remain consistent with Regions 2 and 3 in Figure 1(b). To achieve this low $pO_2$ environment in practice, we react and sinter these oxides in a tube furnace under Ar flow, as illustrated in the experimental setup in Figure 2. Figure 3 presents the X-ray diffraction (XRD) patterns for these Mn- and Fe-containing compositions, compared to the prototypical MgCoNiCuZnO. All samples are sintered at 1100°C for 5 hours: one set in air for comparison (Figure 3(a)) and the other in Ar, as illustrated by the schematic in Figure 2, to validate our hypothesis (Figure 3(b)). Figure 3(a) confirms that all Mn and Fe-containing compositions processed in air predominantly form the spinel structure with a rock salt secondary phase – despite all starting precursors being in their $A^{2+}O^{2-}$ binary oxide form. This underscores Mn and Fe tendencies for higher oxidation states at ambient atmospheric pressure in agreement with the phase diagram in Figure 1(b). In comparison, when we restrict the oxygen availability to that provided by the starting oxide precursors by processing our compositions in Ar, the resulting XRD patterns in Figure 3(b) confirm that all Mn- and Fe-containing compositions predominantly form a single-phase rock salt structure, supporting our arguments and rationale. In other words, flowing only Ar produces $pO_2$ akin to Region 2 in Figure 1(b) (Methods). Combined with the AO starting powders and configurational entropy contributions, this leads to forming single-phase high-entropy rock salts. Notably, the prototypical MgCoNiCuZnO develops a reduced metallic phase after high-temperature processing in Ar, which we attribute to Cu reduction and melting. Figure 3(c) presents the X-ray fluorescence (XRF) spectra for all compositions corresponding to those in Figure 3(b), confirming both their composition and closely equimolar stoichiometry (See Table S1, Supporting Information Note 2 for the details).

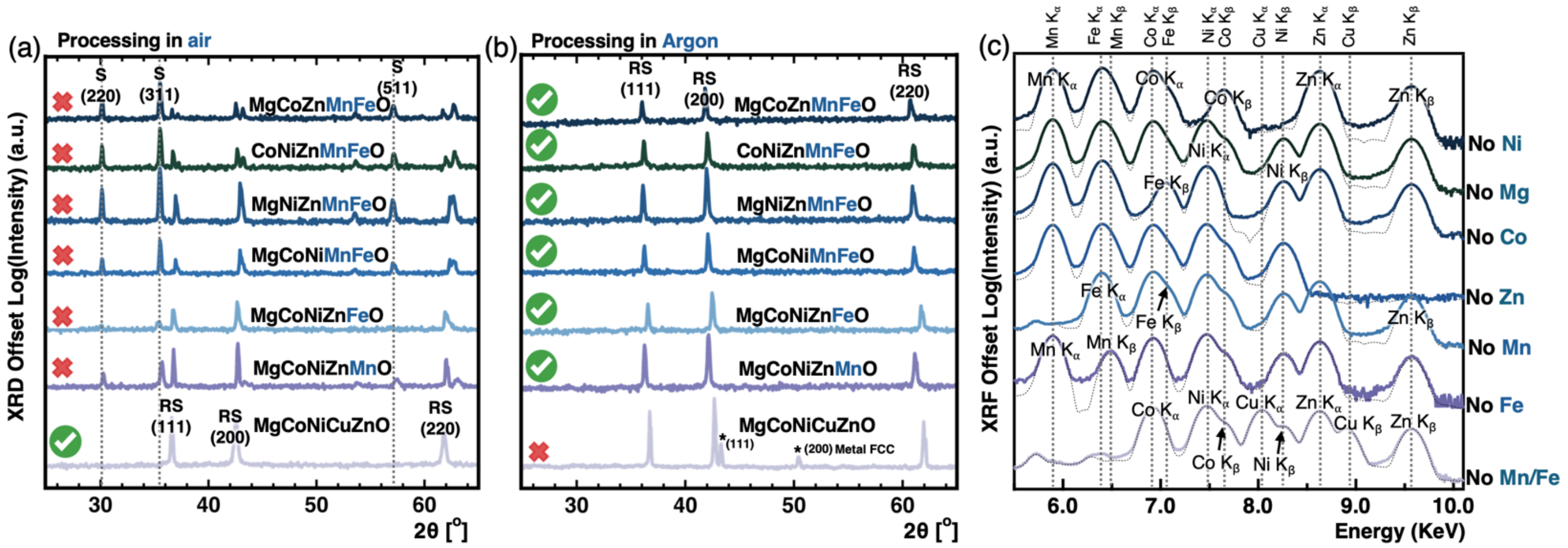

Figure 3: (a) X-ray diffraction scans of prototypical single-phase MgCoNiCuZnO and six five-component systems containing Mn, Fe or both after firing in ambient conditions. All systems containing Mn and Fe form a spinel phase and rock salt secondary phase. (b) X-ray diffraction scans of the same systems under processing in a reducing environment with Ar gas flow. All systems containing Mn and Fe form a single-phase rock salt, while MgCoNiCuZnO forms secondary phases as a result of excess reduction. (c) X-ray fluorescence spectra of each composition highlighting all detected cations except Mg. The exact fitted concentrations are provided in Supporting Information Table S1. 'RS' denotes the peaks corresponding to the rock salt structure, while 'S' denotes those corresponding to the cubic spinel structure.

To confirm that Mn and Fe each maintain a 2+ valence within the HEO matrix, we measure CoNiZnMnFeO and MgNiZnMnFeO X-ray absorption fine structure (XAS) spectra with only the X-ray absorption near edge (XANES) region, both compositions containing Mn, Fe, Ni, Zn, and either Co or Mg. We choose these compositions as they are among the most challenging to stabilize (Figure 1a), as one lacks Co or Mg—key rock salt stabilizers under our synthesis conditions—and both include ZnO, whose wurtzite preference adds structural and enthalpic complexity[3]. The rising absorption edge in Figure 4(a) and 4(c), called the white line, represents electronic transitions that appear as inflection points, which the spectrum's derivative precisely locates. Here, $E_0$, the edge photon energy, is defined as the first main peak in the first derivative (Supporting Information Note 3), excluding contributions from pre-edge features. The $E_0$ values for 3d-transition metal K-edges increases linearly with valence states[17], enabling a best-fit line to estimate unknown valence states as we indicate in Figure 4(b) and 4(d) for Mn and Fe, respectively. Using this approach, all measured cations in MgNiZnMnFeO and CoNiZnMnFeO—including Mn and Fe—predominantly exhibit a 2+ valence state. This is evident from the proximity of their measured $E_0$ edge energies to those corresponding to 2+ from reference samples, as indicated by the orange stars in Figure 4(b) and 4(d).

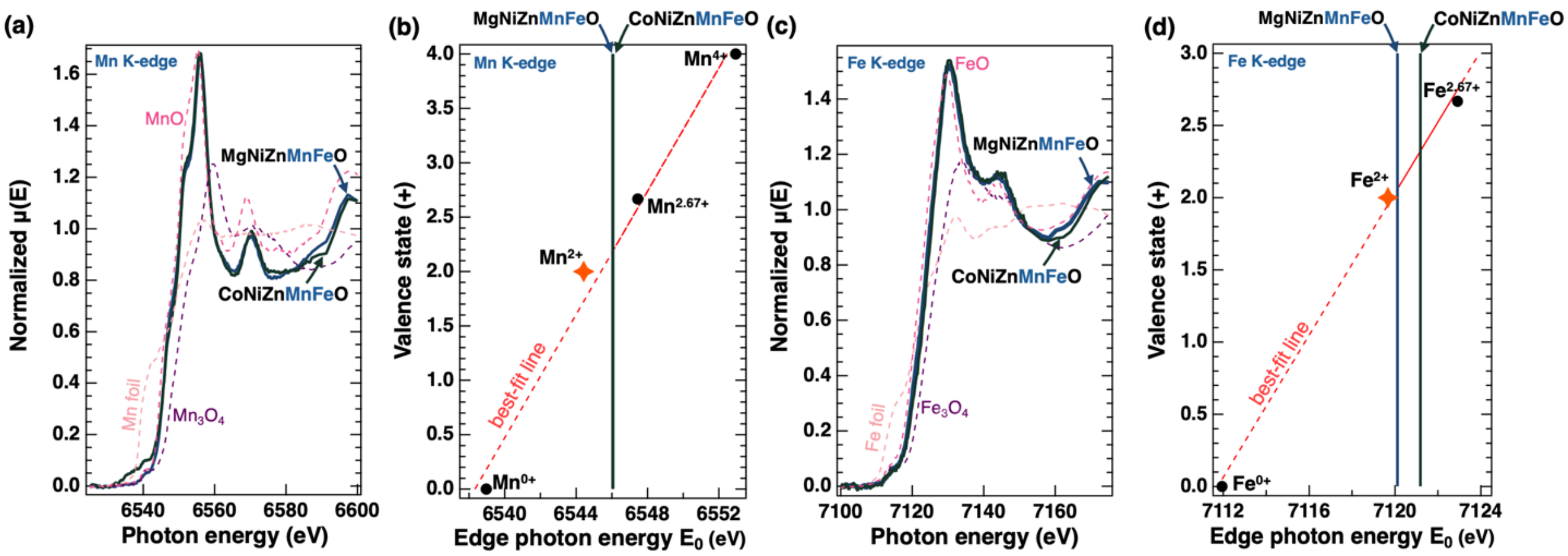

Figure 4: (a) X-ray absorption near edge structure (XANES) measurements of Mn K-edge spectra in MgNiZnMnFeO and CoNiZnMnFeO in comparison to $Mn^{x+}$ reference values. (b) Mn K-edge photon energy versus valence state with best-fit line confirming a predominance of $Mn^{2+}$ within both high entropy compositions (c) XANES measurements of Fe K-edge spectra in MgNiZnMnFeO and CoNiZnMnFeO in comparison to $Fe^{x+}$ reference values. (d) Fe K-edge photon energy vs valence state with best-fit line confirming a predominance of $Fe^{2+}$ within both high entropy compositions. The 2+ reference value is indicated by orange star in (b) and (d).

## *The parent six-component composition: MgCoNiZnMnFeO*

To build on the preceding discussion and further solidify the concepts presented in this work, we propose two key investigation avenues. First, we stabilize the six-component parent composition $Mg_{1/6}Co_{1/6}Ni_{1/6}Zn_{1/6}Mn_{1/6}Fe_{1/6}O$, MgCoNiZnMnFeO for brevity, using the same synthesis conditions that successfully stabilized the six five-component derivative compositions shown in the rock salt HEO phase in Figure 3(b). Second, we test the rock salt phase stability under more stringent reducing conditions during synthesis compared to the reducing conditions in Figure 3(b). Figure 5(a) presents the XRD patterns of MgCoNiZnMnFeO for the same pellet sintered at different temperatures, each for 5 hours under a 100 SCCM Ar flow. Up to 850°C, a rock salt phase predominates, with wurtzite being the only secondary phase which likely corresponds to partially-unreacted ZnO phase. Beyond that temperature, the XRD pattern indicates a single-phase rock-salt structure, further supporting our reasoning and methods (see Supporting Information Note 2

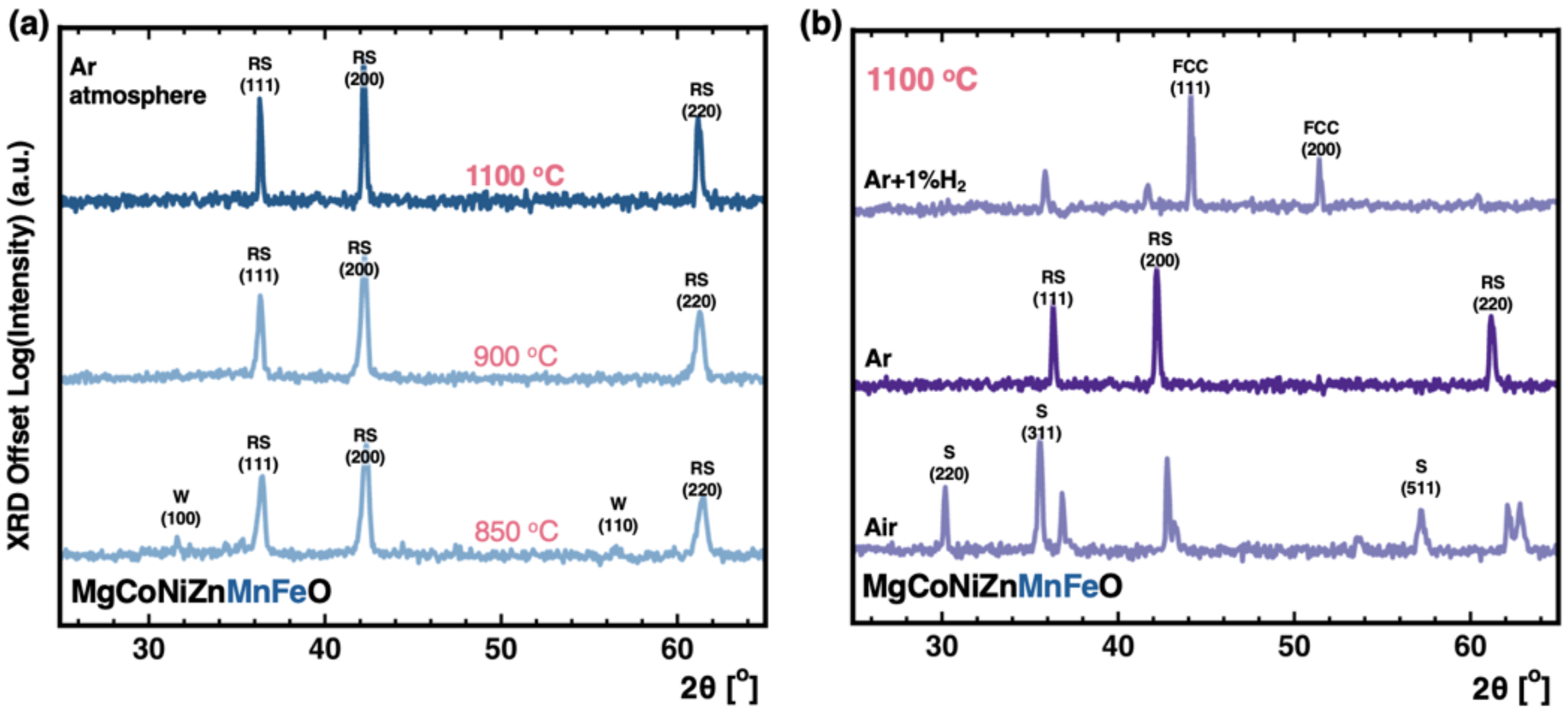

Figure 5: (a) X-ray diffraction patterns of the 6-component high-entropy compositions $Mg_{1/6}Co_{1/6}Ni_{1/6}Zn_{1/6}Mn_{1/6}Fe_{1/6}O$ sintered for 5 hours under 100SCCM of Ar at different temperatures, suggesting that the transition to single phase occurs between 850-900ºC with the disappearance of the wurtzite (W) peaks. (b) X-ray diffraction patterns of $Mg_{1/6}Co_{1/6}Ni_{1/6}Zn_{1/6}Mn_{1/6}Fe_{1/6}O$ sintered at 1100°C under ambient oxygen partial pressure, optimized reducing conditions and excess reducing conditions; while it exhibits single phase rock salt when sintered for 5 hours under 100 SCCM flow of Ar, a metallic phase emerge when either small percentage of $H_2$ is added or if sintered for prolonged time periods. 'RS' denotes peaks from the rock salt structure, 'S' denotes those from the cubic spinel structure, 'W' indicates the wurtzite structure, and 'FCC' indicates the face-centered cubic metallic structure.

for all compositions' wide 2θ–θ scans that we stabilize in single-phase rock salt structure).

In Figure 5(b), we present the second experiment, where three MgCoNiZnMnFeO ceramic pellets are prepared and sintered at 1100 °C for 5 hours, each in a different atmosphere: air, 100SCCM Ar, and 100 SCCM forming gas (99% Ar + 1% $H_2$) (details in Methods). The Ar-sintered sample forms a stable single-phase rock-salt structure, consistent with our previous conclusions, the air-sintered sample is predominantly spinel, and the forming gas sample is partially reduced to metallic phase(s). Notably, an additional 1% $H_2$ lowers $PO_2$ to that in Region 3 in Figure 1(b) or beyond. Therefore, observing those metallic phases follows exactly the valence phase diagram in Figure 1(b), which shows that FeO formation conditions at 1100°C also reduce Co and Ni to their metallic states (see Supporting Information Note 4 for another example demonstrating how metallic phases evolve under forming gas in MgCoNiMnFeO).

## *Chemical potential overlap as a synthesis descriptor*

Finally, we develop a simplistic synthesizability descriptor that both captures the key trends observed in the valence phase diagram (Figure 1b) and can be readily extended to other systems at minimal cost. Although our experimental results thus far focus on relatively high-temperature processing, valuable insights can be gained from ground-state (0 K) density functional theory (DFT) calculations, provided the cation valence states are well represented[3,14,15]. The open-source Materials Project database offers powerful tools for this purpose, including the Phase Diagram and Chemical Potential Diagram modules[18,19]. Developed based on the pioneering work by Yokokawa et al. (Ref.[20]) and detailed by Todd et al. (Ref.[21]), these diagrams map each component's chemical potential onto an axis, delineating different stoichiometries stability regions. When choosing oxygen as a chemical potential axis, the diagram maps different oxidation states on the convex hull. For an arbitrary oxide $A_xO_y$, the y-axis can therefore represent the oxygen chemical potential, while the x-axis corresponds to cation A chemical potential, providing oxygen chemical potential regions where specific cation oxidation states are enthalpically stable. Mg-O, Cu-O and Mn-O chemical potential diagrams are illustrated in Figure 6(a), the diagrams for all 3d transition metals can be found in Supporting Information Note 5. From Figure 6(a), we find an overlap in oxygen chemical potential for $Mg^{2+}$ and $Cu^{2+}$ as well as $Mg^{2+}$ and $Mn^{2+}$, however a separation exists between $Cu^{2+}$ and $Mn^{2+}$ along the oxygen chemical potential axis, in agreement with the valence phase diagram in Figure 1(b) that shows no coexistence between $Cu^{2+}$ and $Mn^{2+}$. We quantify this oxygen chemical potential overlap ($\mu_{overlap}$) for all 21 possible $A_{1/2}A'_{1/2}O$ cation combinations in Figure 6(b), rapidly evaluating common valence stability windows. A larger overlap (indicated in blue shade) suggests a wide stability window where cations coexist in their $A^{2+}O^{2-}$ stoichiometry, making single-phase synthesis feasible provided this window can be accessed experimentally. In contrast, a greater separation (indicated in red shade) suggests a narrower stability window, making synthesis increasingly difficult or even impossible. Notably, this descriptor is impartial to the equilibrium crystal structure as long as it hosts the AO stoichiometry. The values for $\mu_{overlap}$ in Figure 6(b) largely align with the understanding gained from Figure 1(b) such as $Mg^{2+}$ stability across a wide oxygen chemical potential region, while $Cu^{2+}$ with $Mn^{2+}/Fe^{2+}$ mixtures are not stable.

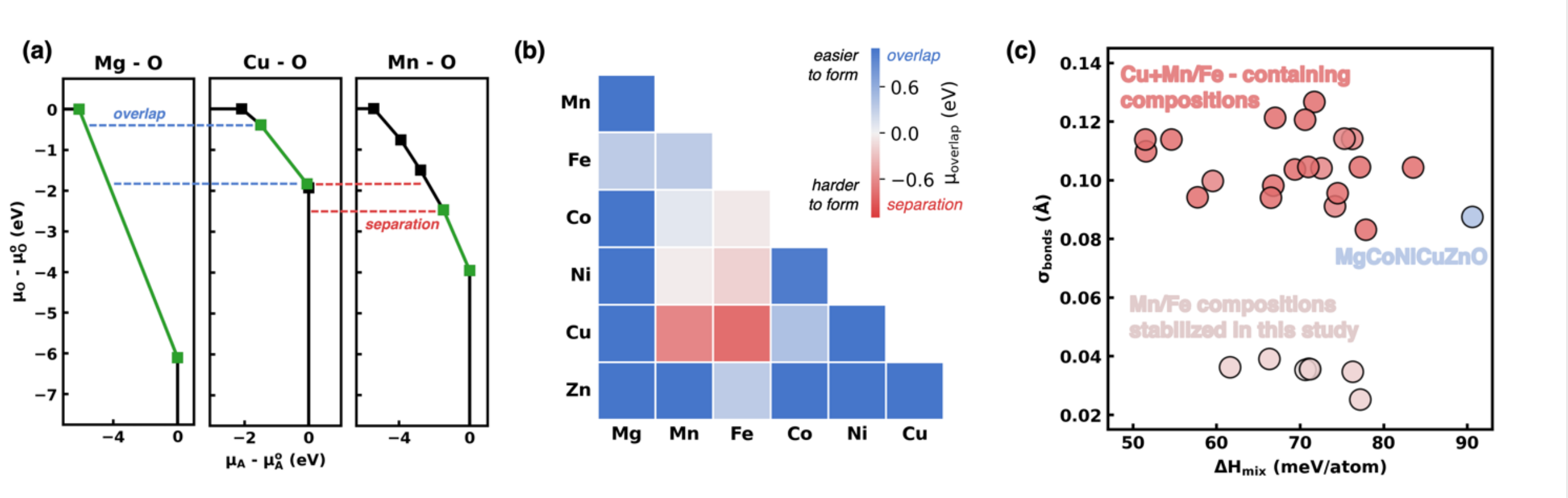

Figure 6: (a) Chemical potential diagrams for Mg-O, Cu-O and Mn-O extracted from the Materials Project database, demonstrating stability windows of each cation in their $A^{2+}O^{2-}$ binary oxides. A green color denotes oxygen chemical potential regions in which $A^{2+}O^{2-}$ compositions are stable. Note the significant overlap MgO has with CuO and MnO, but the separation between CuO and MnO. (b) Chemical potential overlap ($\mu_{overlap}$) for all two-cation AO combinations. Cu, Mn, and Fe consistently have the smallest $\mu_{overlap}$, while Mg and Zn generally have the largest available overlap. (c) Introducing chemical potential overlap as a third descriptor along with $\Delta H_{mix}$ and $\sigma_{bonds}$ reveals the uniqueness of prototypical MgCoNiCuZnO as the only 5- or 6-cation combination with significant overlap in $A^{2+}O^{2-}$ binary oxide stability windows. This diagram also demonstrates the difficulty of stabilizing compositions with Cu, Mn and Fe.

To provide a more comprehensive view for HEO synthesis, we extend the $\mu_{overlap}$ descriptor in Figure 6(c) to all five- and six-cation combinations as a color overlay (Figure 6(c)) in addition to our original descriptors $\Delta H_{mix}$ and $\sigma_{bonds}$ (Figure 1(a)) (details in Methods). We use the same color scheme as in Figure 6(b) to maintain visual consistency. Figure 6(c) suggests that despite MgCoNiCuZnO's relatively high $\Delta H_{mix}$ and $\sigma_{bonds}$ compared to other compositions, it readily forms a single-phase rock salt AO structure as all its cations share an overlapping oxygen chemical potential window. We note that this overlap must coincide with ambient experimental conditions as MgCoNiCuZnO is the only rock salt composition explored that forms a single-phase in air (Figure 3(a)). Conversely, while compositions with Mn and Fe appear easiest to stabilize based on their low $\Delta H_{mix}$ and $\sigma_{bonds}$, the $\mu_{overlap}$ descriptor resolves the greater difficulty to stabilize these compositions compared to MgCoNiCuZnO. Furthermore, compositions containing Cu and Mn/Fe, despite their low $\Delta H_{mix}$, exhibit increased $\sigma_{bonds}$ and a large separation in chemical potential mainly due to Cu and Mn/Fe, making their stabilization more challenging, if not impossible. Hence, for accurate stability predictions, it is critical to consider oxygen chemical potential overlap with conventional descriptors[8,14].While we focus here only on the $A^{2+}O^{2-}$ stoichiometry for rock salt HEOs, this concept will be similarly applicable to other crystal systems.

## Conclusions and outlook

The enthalpic parameters $\Delta H_{mix}$ and $\sigma_{bonds}$ suggest that Mn- and Fe-containing systems should stabilize more easily than the prototypical MgCoNiCuZnO. However, at typical firing temperatures and ambient pressure, Mn and Fe favor the 3+ oxidation state, often forming higher-valence phases like spinels. Introducing oxygen chemical potential overlap ($\mu_{overlap}$) as a third descriptor illuminates MgCoNiCuZnO's uniqueness: despite its higher $\Delta H_{mix}$ and $\sigma_{bonds}$ values, each cation remains in the 2+ state under our synthesis temperature and earth's ambient $pO_2$. In contrast, Mn- and Fe-containing compositions require reduced $pO_2$ to stabilize them in the rock salt phase. By optimizing these conditions, we successfully synthesize seven single-phase HEOs. Expanding equilibrium synthesis descriptors to include oxygen chemical potential broadens the HEO design space, potentially enabling the discovery and realization of new spinel, perovskite, fluorite, bixbyite, and garnet systems. Future studies should also explicitly incorporate finite temperature effect and configurational entropy for improved stability predictions.

While we focus on stabilizing Mn and Fe in the 2+ state, their multivalency offers exciting opportunities, especially under far-from-equilibrium conditions. In our previous studies, we have demonstrated that access to a diverse valence landscape enables control over crystal structure[3,22], microstructure and nanostructure features[23–25], and properties[3,26,27], directly influencing functionality. Mn and Fe addition to the rock salt HEO family likely enables multifunctional material design by facilitating charge trapping, ion migration, and resistive switching while also introducing emergent magnetic interactions[15,28,29]. We therefore anticipate that the compositions synthesized in this work will serve as a foundation for future research exploring magnetism, polaron conduction, and other promising properties through both equilibrium and non-equilibrium synthesis methods.

# Supporting Information

## Methods:

*Bulk Synthesis*

Depending on composition, we combine MgO (Sigma-Aldrich, 342793), CoO (Sigma-Aldrich, 343153), NiO (Sigma-Aldrich, 203882), CuO (Alfa Aesar, 44663), ZnO (Sigma-Aldrich, 96479), MnO (Sigma-Aldrich, 377201), and FeO (Sigma-Aldrich, 400866) in equimolar proportions. We mix and mill the powders with 5 mm yttrium-stabilized zirconia media for 2.5 hours, then press the powder into 2.5 cm pellets at 60-100 MPa (Carver Laboratory Press) for 30 seconds. We fire pellets at 1100 °C in a box furnace under ambient conditions or in a tube furnace (Across International NC2156188) under 100 SCCM *Linde UHP* Ar or 100 SCCM forming gas (99% Ar:1% $H_2$). We quench all samples from ~700 °C to prevent low-temperature phase segregation. When using Ar or forming gas, we quench inside the tube furnace under continuous flow to avoid surface oxide formation. Flowing 100 SCCM Ar at 1100°C in our setup yields $pO_2$ in the range of $10^{-6}$-$10^{-8}$ atm, while flowing 100 SCCM 1%$H_2$ forming gas at 1100°C yields $pO_2$ in the range of $10^{-18}$-$10^{-20}$ atm. We verify structural composition via X-ray diffraction (Panalytical Empyrean) using θ–2θ Bragg-Brentano HD scans with a PIXcel3D detector and identify phases with PANalytical HighScore. We monitor equimolar cation compositions before and after sintering by X-ray fluorescence (Panalytical Epsilon 1).

*X-ray absorption fine structure*

X-ray absorption fine structure (XAFS) spectra are collected using an easyXAFS300+ (Renton, WA) with a Ag X-ray tube operating at 35 kV and 25 mA[30]. Only the X-ray absorption near edge (XANES) region is measured with scans that ranged from 40 eV below to 175 eV above the respective absorption edge. The XAFS samples are massed to an appropriate amount determined via xraydb[31], a database used to calculate masses to use for transmission XAFS samples. These masses are mixed with boron nitride as a filler so that a 5 mm diameter circular pellet could be made. A total of 12 scans are taken per edge per sample, the spectra are then merged and processed using the Demeter package for XAFS analysis[32].

*Thermodynamic Analyses*

Thermodynamic evaluations of the most stable oxidation states are carried out using the OpenCalphad software package[33], employing established thermodynamic data from assessments of relevant binary oxide systems. For each A–O system of interest (A = Mg, Co, Ni, Cu, Zn, Mn, Fe), we retrieve thermodynamic models and parameters for the stable phases from the literature[34–40]. We describe solution phases by the compound energy formalism[41], and we explicitly model stoichiometric compounds as functions of temperature.

We assess phase stabilities in pure $O_2$ environments under an ideal gas assumption, with total gas pressure ranging from $10^{-25}$ bar to $10^5$ bar. To ensure excess oxygen in every calculation, we set the elemental mole fractions $n$(O) = 0.8 and $n$(A) = 0.2. We determine phase-transition temperatures by minimizing the Gibbs free energy at various pressures, which enables us to construct phase boundaries over a range of −25 to 5 atm. We derive cation valences from the nominal stoichiometry of each phase, ignoring small off-stoichiometry effects in vacancy-tolerant solution phases. All calculations use the step-function capability in OpenCalphad[33], systematically exploring temperature–pressure space to identify equilibrium phase assemblies and oxidation states.

*Oxygen Chemical Potential Overlap Descriptor*

Oxygen chemical potential diagrams are constructed using the Materials Project GGA/GGA+U database as a more thorough number of calculations are present compared to the newer $r^2$SCAN database[18]. Our oxygen chemical potential overlap, $\mu_{overlap}$, is defined for an $A_{1/2}B_{1/2}O$ stoichiometry as:

$$\mu_{overlap} = \min(A_2, B_2) - \max(A_1, B_1)$$

where $A_2$ and $B_2$ are the maximum oxygen chemical potential values for a stable $A^{2+}$ and $B^{2+}$ regions, and $A_1$ and $B_1$ are the minimum oxygen chemical potential values for a stable $A^{2+}$ and $B^{2+}$ regions. Note we can extend the $\mu_{overlap}$ definition for a five-component system (ABCDE)O as

$$\mu_{overlap} = \min(A_2, B_2, C_2, D_2, E_2) - \max(A_1, B_1, C_1, D_1, E_1)$$

A positive value for $\mu_{overlap}$ indicates an overlap in oxygen chemical potential space (a wider synthesis window), while a negative value indicates separation (a narrow/impossible synthesis window). Values for $\Delta H_{mix}$ and $\sigma_{bonds}$ descriptors are taken from our previous work[8] calculated using the CHGNet machine learning interatomic potential[13].

**Note 1: $Mn_xFe_yO_\delta$ sintered in air**

In Figure S1, we show that sintering MnO, FeO, or their mixtures at 1100°C in air for 5 hours leads to the expected oxidation to higher valence states. FeO oxidizes to $Fe_2O_3$ adopting the corundum structure, while MnO transforms to $Mn_3O_4$, forming a tetragonal spinel phase. In mixed MnO-FeO samples, additional peaks appear, indicative of bixbyite and cubic spinel structures.

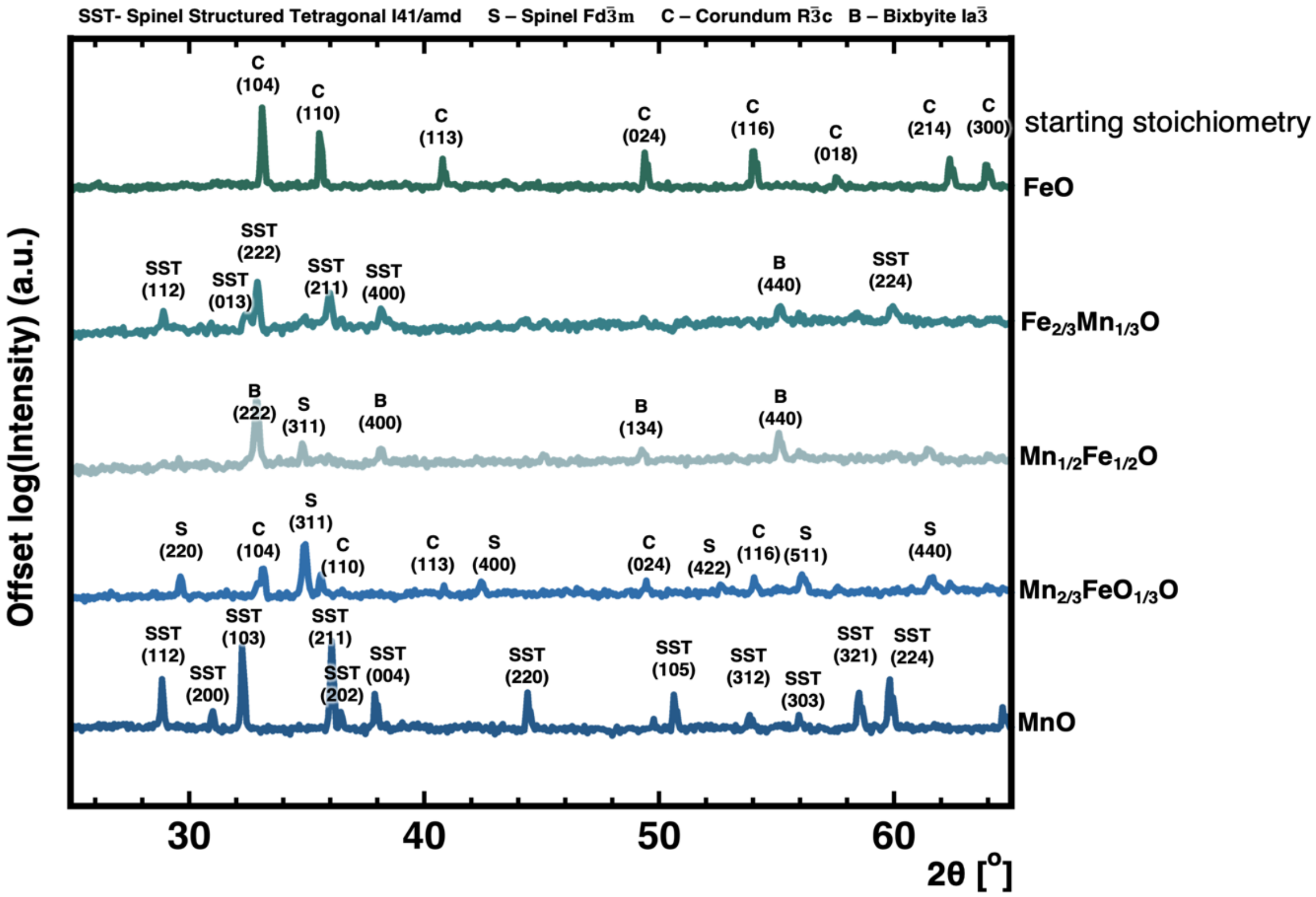

Figure S1. X-ray diffraction (XRD) patterns of MnO, FeO, and MnO-FeO mixtures after sintering at 1100°C in air for 5 hours. Peaks are indexed to their respective crystalline phases.

**Note 2: Wide 2θ- θ scans and XRF results for rock salt high entropy oxide compositions**

Figure S2 presents the XRD patterns of all single-phase rock salt compositions stabilized by carefully controlling the oxygen partial pressure. For reference and completeness, the prototypical MgCoNiCuZnO composition sintered in air is also included. Note that if MgCoNiCuZnO is sintered at 1100ºC under Ar, CuO reduces as discussed in the main manuscript. Table S1 summarizes the X-ray fluorescence (XRF) fitting results, indicating that the compositions are closely equimolar. In our XRF calculations, Mg is fixed at 20% in five-component compositions and 16.667% in six-component compositions, while the remaining cation concentrations are determined from the spectra shown in Figure 3 (main manuscript). We fixed the magnesium concentration because XRF measurements taken in air cannot reliably quantify its content.

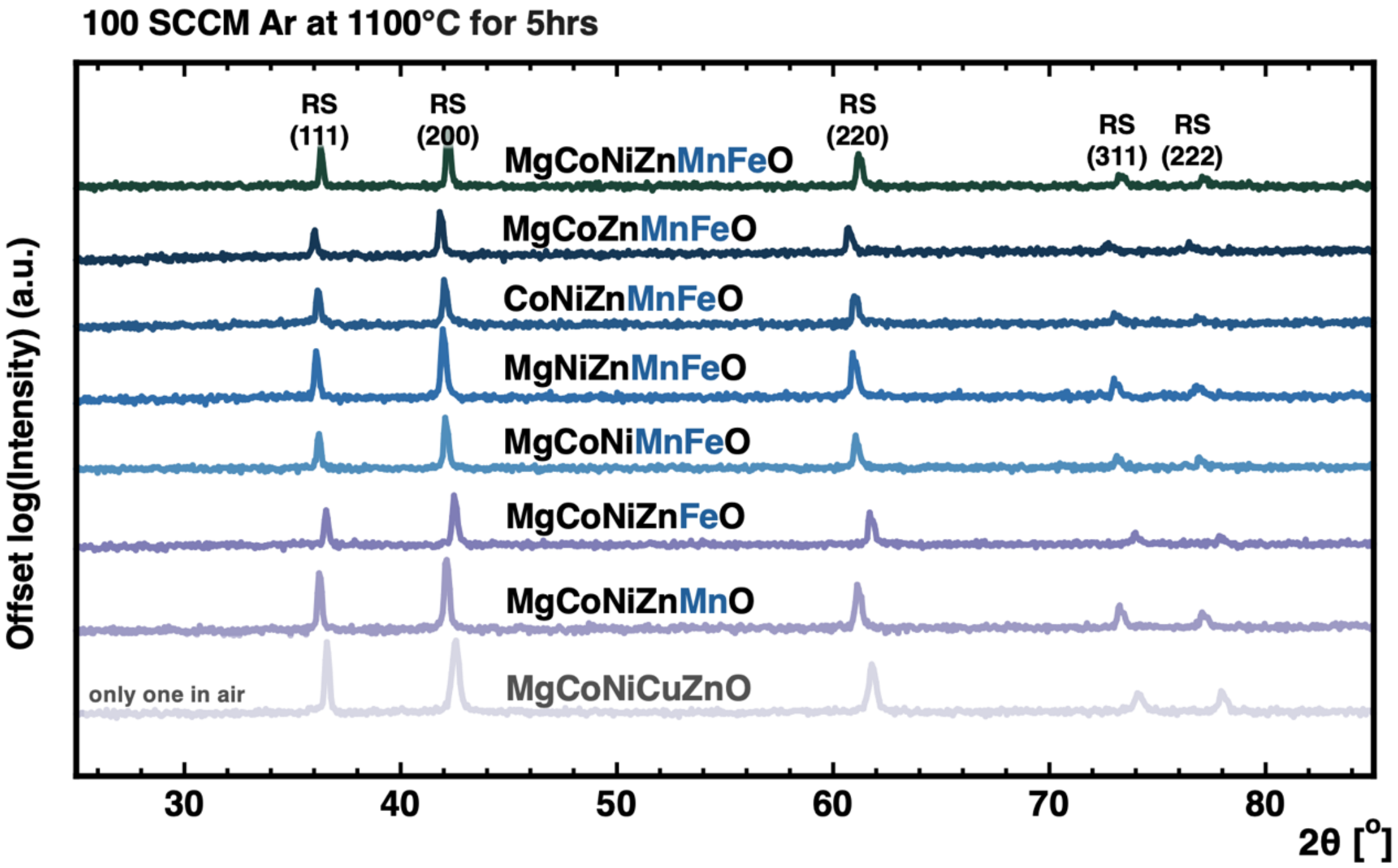

Figure S2. X-ray diffraction patterns of high-entropy compositions containing Mn, Fe, or both, sintered at 1100°C under 100 SCCM of flowing Ar for 5 hours. All compositions exhibit predominantly rock salt structure characteristic peaks. The prototypical MgCoNiCuZnO sintered in air is added as a reference.

Table S1. Percentage cation concentrations measured from XRF spectra.

| Composition | Mg | Co | Ni | Cu | Zn | Mn | Fe |
|---|---|---|---|---|---|---|---|
| MgCoNiZnMnFeO | 16.67 | 17.14 | 17.56 | 0.00 | 16.35 | 15.87 | 16.42 |
| MgCoZnMnFeO | 20.00 | 20.72 | 0.00 | 0.00 | 18.21 | 20.24 | 20.83 |
| CoNiZnMnFeO | 0.00 | 21.24 | 20.90 | 0.00 | 18.40 | 21.14 | 18.33 |
| MgNiZnMnFeO | 20.00 | 0.00 | 21.41 | 0.00 | 16.00 | 21.19 | 21.40 |
| MgCoNiMnFeO | 20.00 | 19.60 | 19.69 | 0.00 | 0.00 | 18.61 | 22.11 |
| MgCoNiZnMnO | 20.00 | 20.53 | 21.07 | 0.00 | 16.84 | 21.56 | 0.00 |
| MgCoNiCuZnO | 20.00 | 20.96 | 19.56 | 18.21 | 21.27 | 0.00 | 0.00 |

## Note 3: XANES spectra and their derivatives

Figures S3 and S4 along with Figure 4 in the manuscript present the full set of collected XANES spectra alongside reference standards, providing the basis for accurate determination of $E_0$. The corresponding derivative spectra are also included to aid in identifying the transition energy. Figure S3 focuses on Co and Ni K-edge data, while Figure S4 shows only the derivative curves for the Mn and Fe K-edges presented in Figure 4. To ensure consistent tracking of binding energies and reliable valence-state assignments, we use the same electronic transition as $E_0$ from the derivative spectra across all samples. Notably, as oxidation state increases, the excitation energy required for the transition also increases.

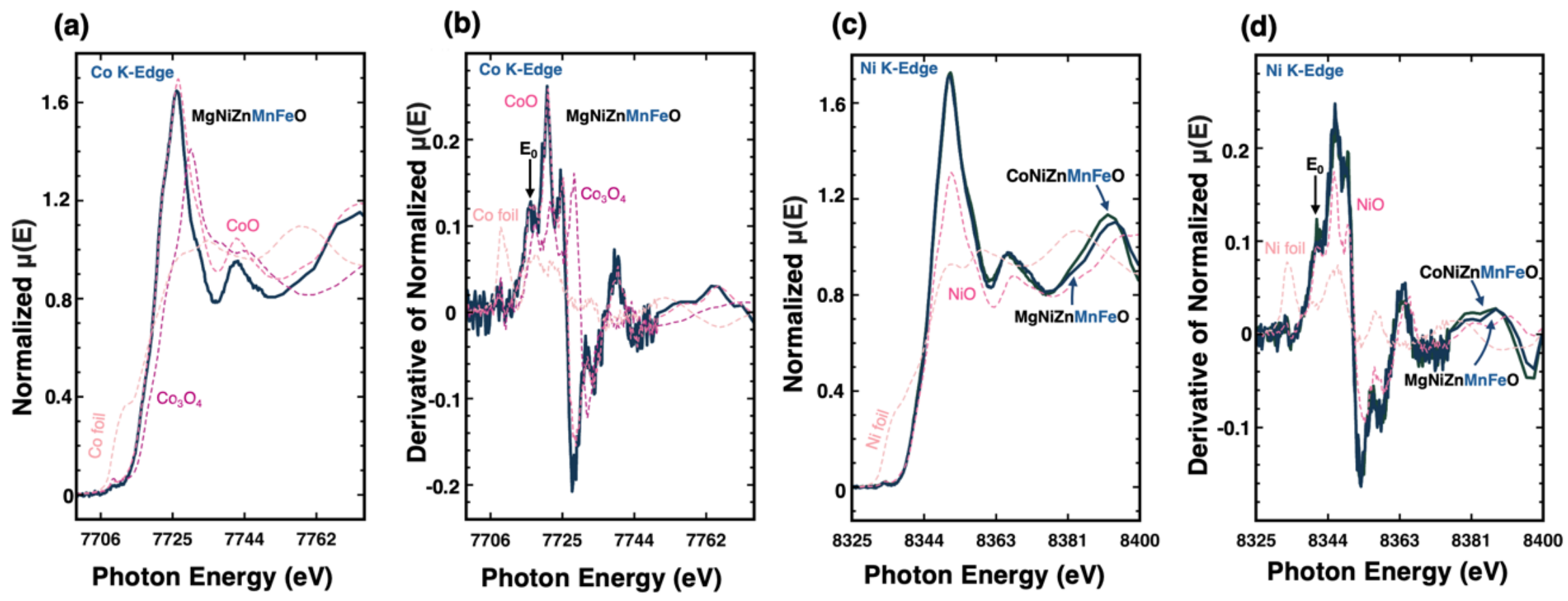

Figure S3. (a) X-ray absorption near edge structure (XANES) spectra of the Co K-edge for MgNiZnMnFeO and CoNiZnMnFeO, compared to reference spectra, with their derivatives shown in (b). (c) XANES spectra at of the Ni K-edge for MgNiZnMnFeO and CoNiZnMnFeO, compared to reference spectra, with the corresponding spectral derivative in (d).

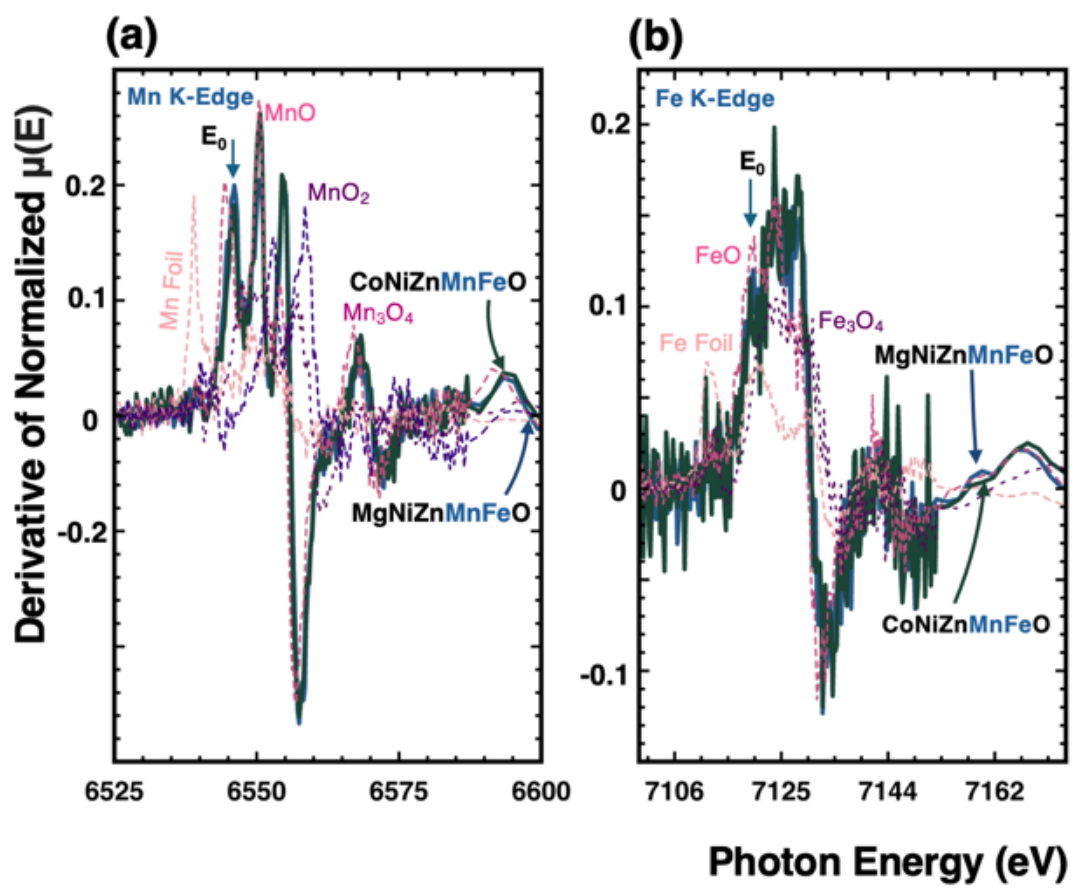

Figure S4. (a) First derivative of the X-ray absorption near edge structure (XANES) spectra at the Mn K-edge for MgNiZnMnFeO and CoNiZnMnFeO, compared to reference spectra, corresponding to Figure 4(a), with $MnO_2$ included as an additional reference. (b) First derivative of the XANES spectra at the Fe K-edge for MgNiZnMnFeO and CoNiZnMnFeO, compared to reference spectra, corresponding to Figure 4(c).

**Note 4: More stringent reducing conditions: MgCoNiMnFeO as another example**

In Figure 5(b) in the main manuscript, we show that more stringent reducing condition by adding 1% $H_2$ to the Ar mixture results in reducing some cations and forming a metallic phase in MgCoNiZnMnFeO. Figure S5 demonstrates this behavior in a five-component composition MgCoNiMnFeO, which is expected across all compositions.

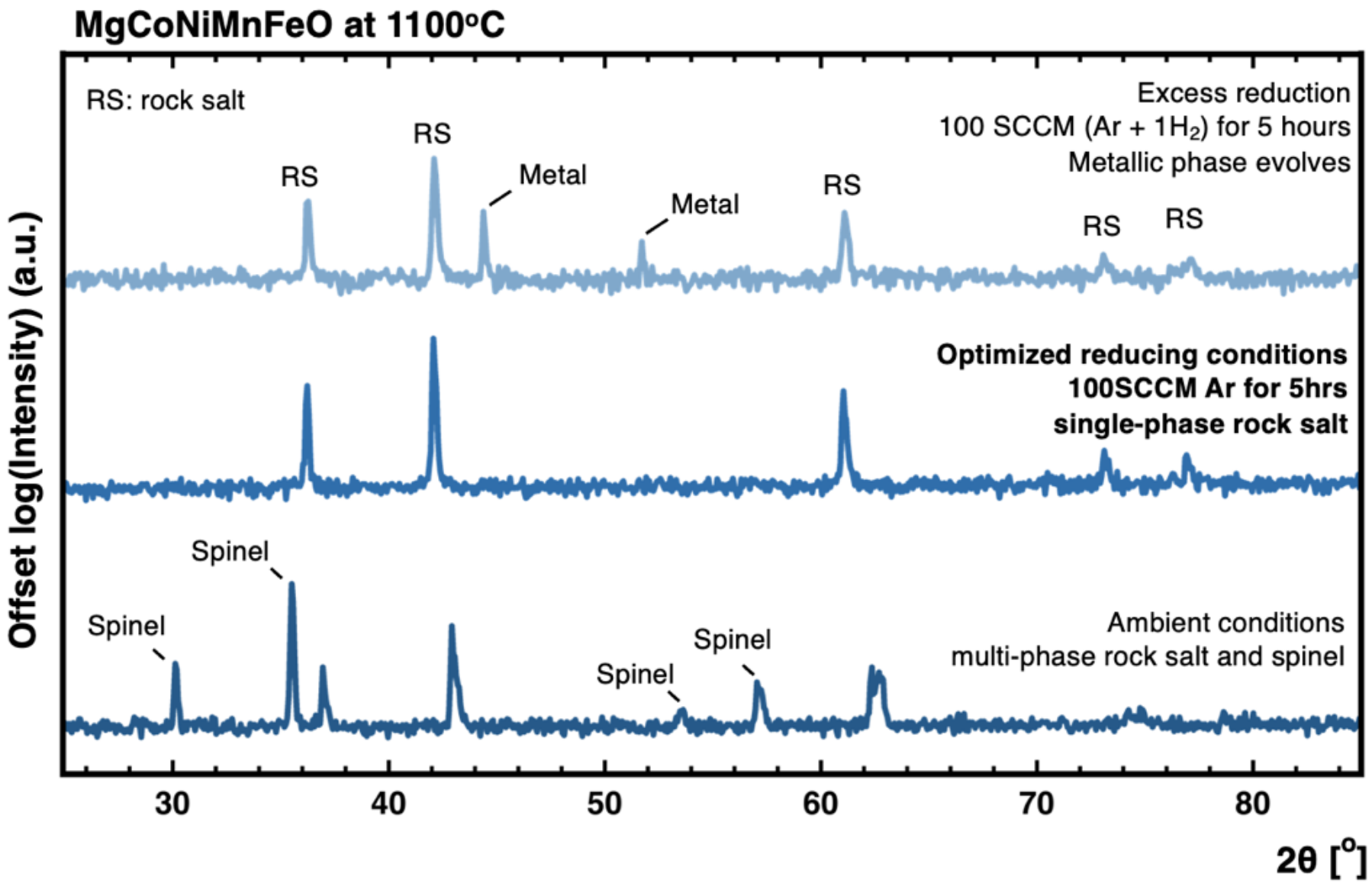

Figure S5. X-ray diffraction patterns $Mg_{1/5}Co_{1/5}Ni_{1/5}Mn_{1/5}Fe_{1/5}O$ sintered at 1100°C under different oxygen partial pressure. Synthesis under forming gas (Ar +1%$H_2$) result in reduced metallic phase, most probably associated with Ni and Co reducing.

## Note 5: Chemical potential diagrams

Figure S6 presents the chemical potentials of 3d transition metals and Mg, extracted from The Materials Project database. The green-bolded region highlights the chemical potential range where $A^{2+}O^{2-}$ is thermodynamically stable. Notably, Sc, V, and Cr do not exhibit a stable $A^{2+}O^{2-}$ phase on the convex hull, whereas Ti forms a stable phase only at very low oxygen chemical potential.

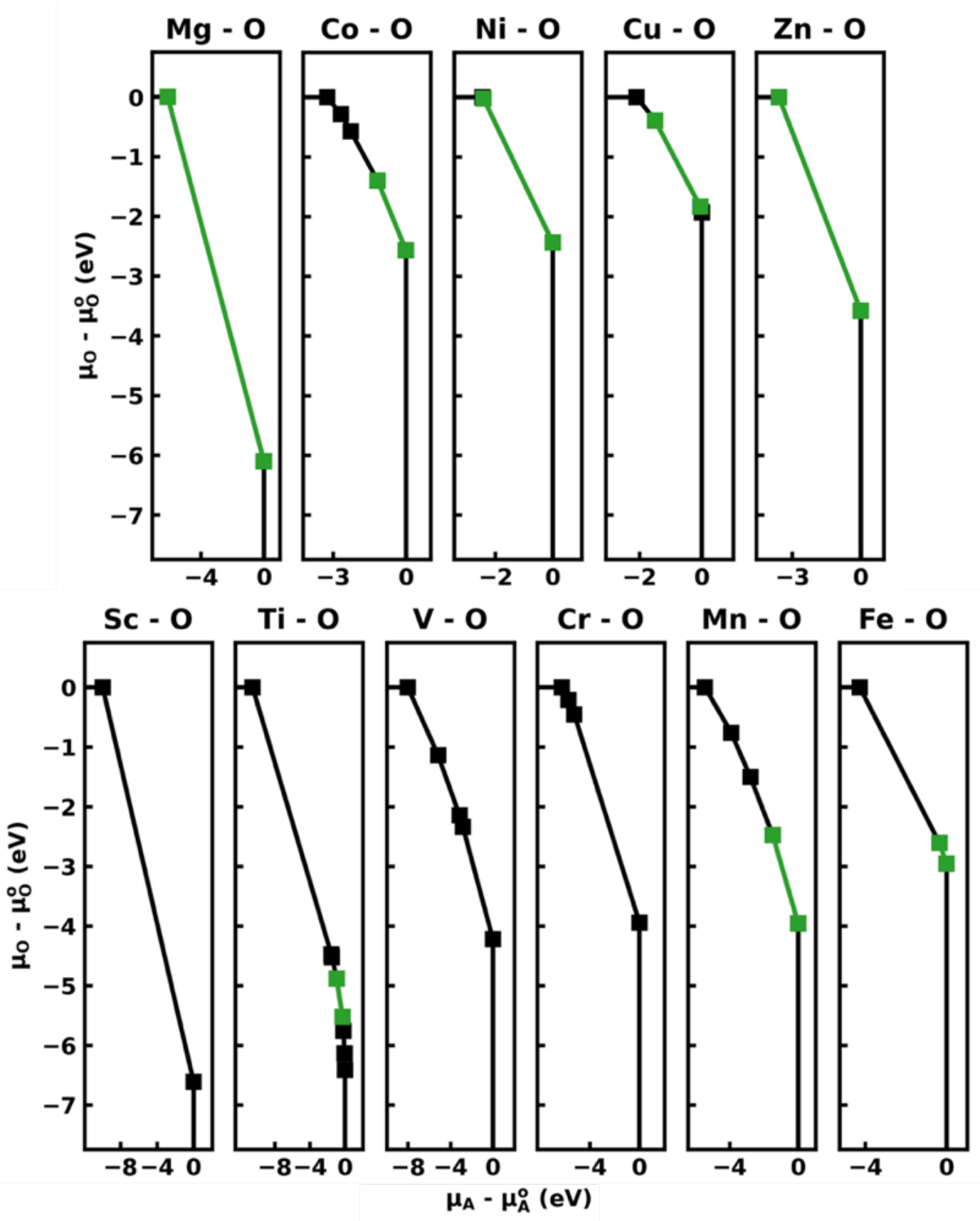

Figure S6. A-O chemical potential diagrams for Mg and all 3d transition metals explored in this study. A green color denotes oxygen chemical potential regions in which $A^{2+}O^{2-}$ compositions are stable.

## Note 6: Chemical potential overlap descriptor

In Table S2, we summarize the values for the computational descriptors for equimolar HEO compositions from the cation cohort: Mg, Mn, Fe, Co, Ni, Cu, and Zn. Compositions are sorted by the oxygen chemical potential overlap descriptor proposed in this work, $\mu_{overlap}$. The refined single-phase stability thresholds are defined from our previous work[8] to determine predicted single-phase stability thresholds ($\Delta H_{mix}$ = 92.2 meV/atom, $\sigma_{bonds}$ = 0.102 Å) as well as $\mu_{overlap}$ = -0.191 eV. Predicted and experimental stability are indicated as single- and multi-phase with green circles and red crosses, respectively. Compositions not yet explored experimentally are indicated by two dashes. For clarity, a red color is used to denote the descriptor(s) values that lead to multi-phase predictions. A complete set of values, including those for Ca, is provided in our previous work (see Ref.[8]).

Table S2. Computational descriptors for equimolar HEO compositions

| Composition | $\mu_{overlap}$ (eV) | $\Delta H_{mix}$ (eV/atom) | $\sigma_{bonds}$ (Å) | Pred. Stability | Expt. Stability |
|---|---|---|---|---|---|
| $Mg_{1/4}Ni_{1/4}Cu_{1/4}Zn_{1/4}O$ | 1.443 | 0.097 | 0.102 | ✖ | ✖ [3] |
| $Mg_{1/4}Co_{1/4}Ni_{1/4}Zn_{1/4}O$ | 1.035 | 0.079 | 0.020 | ● | ● [3] |
| $Mg_{1/4}Co_{1/4}Ni_{1/4}Cu_{1/4}O$ | 0.436 | 0.077 | 0.099 | ● | ● [3] |
| $Mg_{1/4}Co_{1/4}Cu_{1/4}Zn_{1/4}O$ | 0.436 | 0.093 | 0.116 | ✖ | ✖ [3] |
| $Co_{1/4}Ni_{1/4}Cu_{1/4}Zn_{1/4}O$ | 0.436 | 0.100 | 0.116 | ✖ | ✖ [3] |
| $Mg_{1/5}Co_{1/5}Ni_{1/5}Cu_{1/5}Zn_{1/5}O$ | 0.436 | 0.091 | 0.087 | ● | ● [1] |
| $Mg_{1/4}Mn_{1/4}Fe_{1/4}Zn_{1/4}O$ | 0.336 | 0.070 | 0.042 | ● | -- |
| $Mg_{1/4}Mn_{1/4}Co_{1/4}Zn_{1/4}O$ | 0.091 | 0.072 | 0.037 | ● | -- |
| $Mg_{1/4}Mn_{1/4}Co_{1/4}Ni_{1/4}O$ | -0.041 | 0.060 | 0.036 | ● | -- |
| $Mg_{1/4}Mn_{1/4}Ni_{1/4}Zn_{1/4}O$ | -0.041 | 0.061 | 0.039 | ● | -- |
| $Mn_{1/4}Co_{1/4}Ni_{1/4}Zn_{1/4}O$ | -0.041 | 0.079 | 0.036 | ● | -- |
| $Mg_{1/5}Mn_{1/5}Co_{1/5}Ni_{1/5}Zn_{1/5}O$ | -0.041 | 0.071 | 0.036 | ● | ● [15] |
| $Mg_{1/4}Mn_{1/4}Fe_{1/4}Co_{1/4}O$ | -0.059 | 0.057 | 0.036 | ● | -- |
| $Mg_{1/4}Fe_{1/4}Co_{1/4}Zn_{1/4}O$ | -0.059 | 0.075 | 0.024 | ● | -- |
| $Mn_{1/4}Fe_{1/4}Co_{1/4}Zn_{1/4}O$ | -0.059 | 0.078 | 0.030 | ● | -- |
| $Mg_{1/5}Mn_{1/5}Fe_{1/5}Co_{1/5}Zn_{1/5}O$ | -0.059 | 0.071 | 0.035 | ● | ● |
| $Mg_{1/4}Mn_{1/4}Fe_{1/4}Ni_{1/4}O$ | -0.191 | 0.053 | 0.042 | ● | -- |
| $Mg_{1/4}Fe_{1/4}Co_{1/4}Ni_{1/4}O$ | -0.191 | 0.071 | 0.026 | ● | -- |
| $Mg_{1/4}Fe_{1/4}Ni_{1/4}Zn_{1/4}O$ | -0.191 | 0.068 | 0.027 | ● | -- |
| $Mn_{1/4}Fe_{1/4}Co_{1/4}Ni_{1/4}O$ | -0.191 | 0.059 | 0.036 | ● | -- |
| $Mn_{1/4}Fe_{1/4}Ni_{1/4}Zn_{1/4}O$ | -0.191 | 0.076 | 0.039 | ● | -- |
| $Fe_{1/4}Co_{1/4}Ni_{1/4}Zn_{1/4}O$ | -0.191 | 0.090 | 0.025 | ● | -- |
| $Mg_{1/5}Mn_{1/5}Fe_{1/5}Co_{1/5}Ni_{1/5}O$ | -0.191 | 0.062 | 0.036 | ● | ● |
| $Mg_{1/5}Mn_{1/5}Fe_{1/5}Ni_{1/5}Zn_{1/5}O$ | -0.191 | 0.066 | 0.039 | ● | ● |
| $Mg_{1/5}Fe_{1/5}Co_{1/5}Ni_{1/5}Zn_{1/5}O$ | -0.191 | 0.077 | 0.025 | ● | ● |
| $Mn_{1/5}Fe_{1/5}Co_{1/5}Ni_{1/5}Zn_{1/5}O$ | -0.191 | 0.076 | 0.035 | ● | ● |
| $Mg_{1/6}Mn_{1/6}Fe_{1/6}Co_{1/6}Ni_{1/6}Zn_{1/6}O$ | -0.191 | 0.071 | 0.036 | ● | ● |
| $Mg_{1/4}Mn_{1/4}Co_{1/4}Cu_{1/4}O$ | -0.640 | 0.057 | 0.128 | ✖ | -- |
| $Mg_{1/4}Mn_{1/4}Ni_{1/4}Cu_{1/4}O$ | -0.640 | 0.053 | 0.115 | ✖ | -- |
| $Mg_{1/4}Mn_{1/4}Cu_{1/4}Zn_{1/4}O$ | -0.640 | 0.075 | 0.137 | ✖ | -- |
| $Mn_{1/4}Co_{1/4}Ni_{1/4}Cu_{1/4}O$ | -0.640 | 0.050 | 0.126 | ✖ | -- |
| $Mn_{1/4}Co_{1/4}Cu_{1/4}Zn_{1/4}O$ | -0.640 | 0.080 | 0.149 | ✖ | -- |
| $Mn_{1/4}Ni_{1/4}Cu_{1/4}Zn_{1/4}O$ | -0.640 | 0.079 | 0.136 | ✖ | -- |
| $Mg_{1/5}Mn_{1/5}Co_{1/5}Ni_{1/5}Cu_{1/5}O$ | -0.640 | 0.060 | 0.100 | ✖ | -- |
| $Mg_{1/5}Mn_{1/5}Co_{1/5}Cu_{1/5}Zn_{1/5}O$ | -0.640 | 0.075 | 0.114 | ✖ | -- |
| $Mg_{1/5}Mn_{1/5}Ni_{1/5}Cu_{1/5}Zn_{1/5}O$ | -0.640 | 0.073 | 0.104 | ✖ | -- |
| $Mn_{1/5}Co_{1/5}Ni_{1/5}Cu_{1/5}Zn_{1/5}O$ | -0.640 | 0.076 | 0.114 | ✖ | -- |
| $Mg_{1/6}Mn_{1/6}Co_{1/6}Ni_{1/6}Cu_{1/6}Zn_{1/6}O$ | -0.640 | 0.074 | 0.091 | ✖ | -- |
| $Mg_{1/4}Mn_{1/4}Fe_{1/4}Cu_{1/4}O$ | -0.790 | 0.048 | 0.135 | ✖ | -- |
| $Mg_{1/4}Fe_{1/4}Co_{1/4}Cu_{1/4}O$ | -0.790 | 0.063 | 0.119 | ✖ | -- |
| $Mg_{1/4}Fe_{1/4}Ni_{1/4}Cu_{1/4}O$ | -0.790 | 0.060 | 0.107 | ✖ | -- |
| $Mg_{1/4}Fe_{1/4}Cu_{1/4}Zn_{1/4}O$ | -0.790 | 0.073 | 0.127 | ✖ | -- |
| $Mn_{1/4}Fe_{1/4}Co_{1/4}Cu_{1/4}O$ | -0.790 | 0.047 | 0.139 | ✖ | -- |
| $Mn_{1/4}Fe_{1/4}Ni_{1/4}Cu_{1/4}O$ | -0.790 | 0.045 | 0.135 | ✖ | -- |
| $Mn_{1/4}Fe_{1/4}Cu_{1/4}Zn_{1/4}O$ | -0.790 | 0.073 | 0.157 | ✖ | -- |
| $Fe_{1/4}Co_{1/4}Ni_{1/4}Cu_{1/4}O$ | -0.790 | 0.062 | 0.118 | ✖ | -- |
| $Fe_{1/4}Co_{1/4}Cu_{1/4}Zn_{1/4}O$ | -0.790 | 0.087 | 0.135 | ✖ | -- |
| $Fe_{1/4}Ni_{1/4}Cu_{1/4}Zn_{1/4}O$ | -0.790 | 0.087 | 0.125 | ✖ | -- |
| $Mg_{1/5}Mn_{1/5}Fe_{1/5}Co_{1/5}Cu_{1/5}O$ | -0.790 | 0.055 | 0.114 | ✖ | -- |
| $Mg_{1/5}Mn_{1/5}Fe_{1/5}Ni_{1/5}Cu_{1/5}O$ | -0.790 | 0.052 | 0.110 | ✖ | -- |
| $Mg_{1/5}Mn_{1/5}Fe_{1/5}Cu_{1/5}Zn_{1/5}O$ | -0.790 | 0.067 | 0.121 | ✖ | -- |
| $Mg_{1/5}Fe_{1/5}Co_{1/5}Ni_{1/5}Cu_{1/5}O$ | -0.790 | 0.067 | 0.094 | ✖ | -- |
| $Mg_{1/5}Fe_{1/5}Co_{1/5}Cu_{1/5}Zn_{1/5}O$ | -0.790 | 0.077 | 0.104 | ✖ | -- |
| $Mg_{1/5}Fe_{1/5}Ni_{1/5}Cu_{1/5}Zn_{1/5}O$ | -0.790 | 0.075 | 0.096 | ✖ | -- |
| $Mn_{1/5}Fe_{1/5}Co_{1/5}Ni_{1/5}Cu_{1/5}O$ | -0.790 | 0.051 | 0.114 | ✖ | -- |
| $Mn_{1/5}Fe_{1/5}Co_{1/5}Cu_{1/5}Zn_{1/5}O$ | -0.790 | 0.072 | 0.127 | ✖ | -- |
| $Mn_{1/5}Fe_{1/5}Ni_{1/5}Cu_{1/5}Zn_{1/5}O$ | -0.790 | 0.071 | 0.121 | ✖ | -- |
| $Fe_{1/5}Co_{1/5}Ni_{1/5}Cu_{1/5}Zn_{1/5}O$ | -0.790 | 0.084 | 0.104 | ✖ | -- |
| $Mg_{1/6}Mn_{1/6}Fe_{1/6}Co_{1/6}Ni_{1/6}Cu_{1/6}O$ | -0.790 | 0.058 | 0.094 | ✖ | -- |
| $Mg_{1/6}Mn_{1/6}Fe_{1/6}Co_{1/6}Cu_{1/6}Zn_{1/6}O$ | -0.790 | 0.069 | 0.104 | ✖ | -- |
| $Mg_{1/6}Mn_{1/6}Fe_{1/6}Ni_{1/6}Cu_{1/6}Zn_{1/6}O$ | -0.790 | 0.067 | 0.098 | ✖ | -- |
| $Mg_{1/6}Fe_{1/6}Co_{1/6}Ni_{1/6}Cu_{1/6}Zn_{1/6}O$ | -0.790 | 0.078 | 0.083 | ✖ | -- |
| $Mn_{1/6}Fe_{1/6}Co_{1/6}Ni_{1/6}Cu_{1/6}Zn_{1/6}O$ | -0.790 | 0.071 | 0.104 | ✖ | -- |